\documentclass[12pt]{article}
\usepackage{graphicx}
\usepackage[top=3cm, bottom=2cm, left=3cm, right=3cm]{geometry}
\title{Cosmic ray anisotropies to 5 PeV}
\author{A. D. Erlykin $^{1,2}$ and A. W. Wolfendale $^{2}$\\
$(1)$ P N Lebedev Physical Institute, Moscow, Russia\\
$(2)$ Department of Physics, Durham University, Durham, UK}
\begin{document}
\maketitle

\begin{abstract}
Several large cosmic ray (CR) detectors have recently provided data on the arrival 
directions of CR, which taken together with previous data recorded over many decades 
allow the amplitude and phase of the first harmonic to be derived with reasonable 
precision and up to higher energies. We find a high degree of consistency amongst the 
various measurements. The new data indicate that at an energy above 
$\sim$0.1 PeV a change of the CR anisotropy sets in. The amplitude of the first 
harmonic, which rises to 3 TeV, then diminishes and begins to rise again. The direction
 of the phase also changes to the 
opposite one. A measure of understanding follows from the use of two-dimensional maps 
of cosmic ray excesses over the mean background. When the energy of cosmic rays 
approaches the PeV region, the excess of cosmic rays moves from the Galactic 
Anti-Centre to the opposite direction of the Galactic Centre. The possible role of such
  potential cosmic ray sources as the supernovae Monogem Ring and Vela, which could 
help to explain some of the observed results, is discussed.
\end{abstract}
\section{Introduction}
It is very well known that the biggest problem in the quest to find the origin of CR is
 the presence of magnetic fields in the Interstellar Medium (ISM). The fields have both
 regular and irregular
 components and both play a role in causing the CR to travel by torturous paths. The 
irregular component, which dominates, causes the CR to effectively diffuse from their 
sources.

The present work comprises a quick survey of the results on CR anisotropies which, 
hopefully, have relevance to the origin problem, together with the necessary 
information
about the mode of transport of the CR, specifically the regular and irregular field 
properties. Particular emphasis is given to the new data at sub-PeV and PeV energies 
which appeared since our latest paper on this subject \cite{EW1}.  

A particular quest is to see if there is evidence favouring our 'Single Source Model' 
(SSM) (~see \cite{EW2} and later papers~), in which a particular supernova remnant 
(SNR) (~presumed to be 
the Monogem Ring~) is the main cause of the sharp knee in the CR energy spectrum at 
about 3-4 PeV. According to our model a change in anisotropy and phase might be 
expected at energies approaching 1 PeV. A similar scenario has been invoked by 
Sveshnikova et al. (2011) \cite{Svesh} in terms of one or more local SNR being 
responsible for some, at least, of the 'fine structure' of CR energy spectra and 
the form of the declination ($\delta$) versus right ascention ($RA$) plots. Also 
included in the analysis is a study of the anisotropy of the lower energy CR, i.e. from
 $logE = 2$ to $logE = 5$, where $E$ is in GeV.   
\section{The measured cosmic ray anisotropies}
\subsection{Hazards in measuring the anisotropies} 
In principle, the CR intensity will be a complicated function of Galactic longitude and
 latitude ($\ell, b$) or, equivalently, right ascention and declination ($RA, \delta$).
 It is customary, and reasonable, however, to represent it by two components. The 
first is in terms of a simple flow - with a near-sinusoidal intensity variation, in a 
fixed declination band (~insofar as the detectors sweep out a band of constant 
declination~). This gives the anisotropy amplitude $A$, and the phase in $RA$. The
second is the residual excess, or excesses, which remain after the first harmonic has 
been subtracted. 

When only a limited declination range is available, the measured value of $A$ 
has to be corrected to allow for the fact that the declination at which the CR maximum 
is detected is not necessarily the centre of the declination range available for study.
 This point was made by Kiraly et al. (1979) \cite{Kira} and very recently by 
Lidvansky et al. (2012) \cite{Lidv}. 

Even in the past, when only measurements of the counting rate (~as, for example, with 
 neutron monitors~)
were used for the study of anisotropy, or the arrival directions of CR particles were 
measured with a large uncertainty (~muon telescopes~), it was noted that the use of 
just the first harmonic was not enough to get a good fit of the CR spatial variation. 
The use of two harmonics improves the fit, but it means that the spatial distribution 
of the CR intensity has a rather complicated structure. Nowadays, when the arrival 
directions of the CR particles are measured with good accuracy, the constructed 
two-dimensional maps of the CR intensity have confirmed the complicated structure of 
the CR sky and allowed its analysis on different angular scales.    

\subsection{Summary of the amplitudes and phases of the first harmonics}
Figure 1 shows a summary of the estimates of amplitude and phase derived from a number 
of previous summaries, specifically \cite{EW1}(~to be referred to as EW1~). This new  
summary includes the latest data from Super-Kamiokande \cite{Guil}, 
MILAGRO \cite{Abdo}, ARGO-YBJ \cite{Argo}, IceCube \cite{Icub} and IceTop \cite{Itop}, 
as well as updated data from EAS-TOP \cite{EasT}, which were not shown in our previous 
summary EW1. The errors of the amplitudes and phases of these latest works are shown as
 they are given by the authors, except for IceTop, about which we shall comment later. 
\begin{center}
\begin{figure}[ht]
\includegraphics[height=15cm,width=10.5cm,angle=-90]{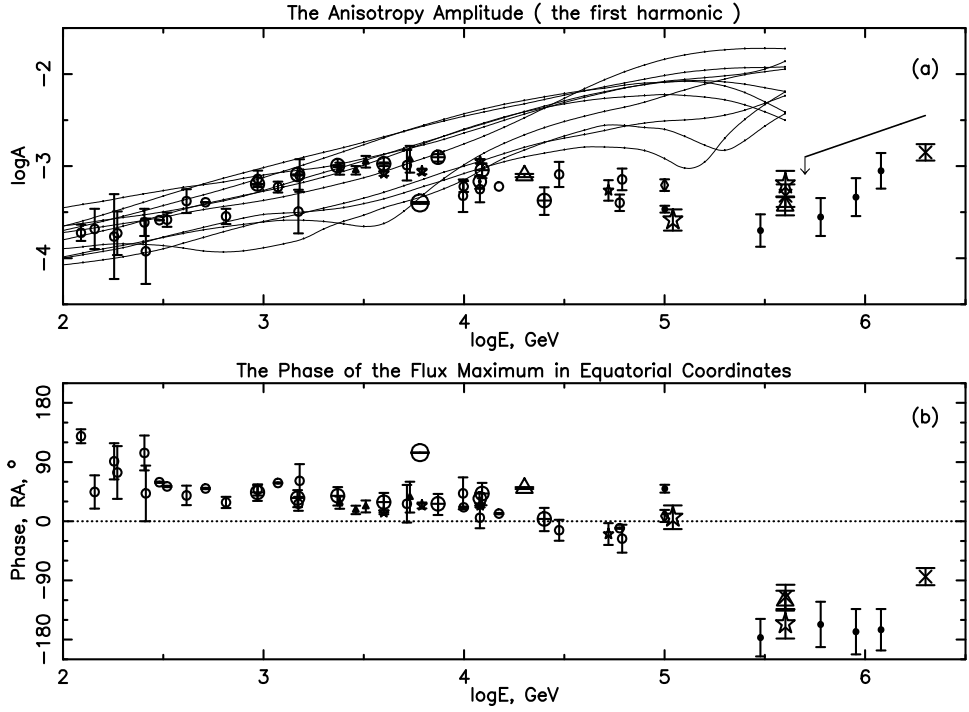}
\caption{\footnotesize The observed amplitude (a) and the equatorial phase (b) of the 
first harmonic of the CR anisotropy. The bulk of the data were taken from Figure 5 of 
our previous survey EW1. The new data are denoted as $\odot$ - Super-Kamiokande 
\cite{Guil}, $\bigcirc$ - MILAGRO \cite{Abdo}, $\oplus$ - ARGO-YBJ \cite{Argo}, $\star$
 - updated EASTOP \cite{EasT}, $\bigtriangleup$ - IceCube \cite{Icub},  $\times$ - 
IceTop \cite{Itop}. Thin full lines - the calculations for 10 different configurations
 of SNR including a single source which is tentatively associated with the Monogem Ring
 SNR \cite{EW1}.} 
\label{fig:fig1}
\end{figure}
\end{center}
The IceCube measurements cover the energy range from tens to hundreds of TeV 
\cite{Icub}. The distribution of primary energies responsible for the production of the
 detected TeV muons is rather wide. The median energies for two selected energy bands 
are 0.02 and 0.4 PeV. It can be remarked that in this energy range the amplitude of the
 first harmonic
 falls down from $(0.79\pm 0.01_{stat}\pm 0.03_{syst})\cdot 10^{-3}$ to 
$(0.37\pm 0.07_{stat}\pm 0.07_{syst})\cdot 10^{-3}$ and the phase changes from 
$50.5\pm 1.0_{stat}\pm 1.1_{syst}$ degrees to $-120.8\pm 10.6_{stat}\pm 10.8_{syst}$ 
degrees. 

The IceTop measurements cover the range of higher energies \cite{Itop}. The median 
values rise from 0.4 PeV to 2 PeV and at the lower energy band they overlap with 
IceCube data giving the possibility to make a comparison between two independent 
experiments carried out at the same location at the South Pole and covering the same 
primary energies. However, a direct comparison of the published results is 
not possible. The IceTop collaboration prefers to fit their intensity profile 
by the superposition of a flat background and a 'gaussian' dip. In order to
 make the comparison with IceCube and other works we fitted the IceTop profile 
with the traditional two-harmonic function. The result of the best fit for both
 0.4 PeV and 2 PeV energy domains is given in Table 1 and shown in Figure 1 for the 
first harmonic.
\begin{center}
Table 1 \\
{\footnotesize Anisotropy amplitude and phase for two energy domains in IceTop}
\begin{tabular}{||c|c|c|c|c|c||} \hline \hline
E$^{med}$,PeV&$A_1(10^{-3})$&RA$_1$(deg)&$A_2(10^{-3})$&RA$_2$(deg)&$\chi^2/ndf$ \\ \hline
0.4 &$0.48\pm0.16$&$245\pm19$&$0.39\pm0.15$&$0.72\pm1.12$&26.5/14 \\ \hline 
2.0 &$1.41\pm0.32$&$276\pm13$&$-0.69\pm0.32$&$257\pm13$&7.1/14 \\ \hline \hline
\end{tabular}    
\end{center}
It is seen that in the 0.4 PeV energy band, where both IceCube and IceTop measurements 
overlap, there is good agreement between the amplitudes and phases of the anisotropy in
 both experiments, which gives some measure of credibility for their results at lower 
and higher energies. IceTop data indicate that in the PeV region the amplitude of the 
first harmonic starts to rise and also the phase is changing.  

Recent measurements of the anisotropy in the Northern Hemisphere were made by 
Super-Kamiokande \cite{Guil}, MILAGRO \cite{Abdo}, ARGO-YBJ \cite{Argo} and the 
updated EAS-TOP data \cite{EasT}. They did not reach the energy region above 0.1 
PeV, but are useful because they are complementary to the measurements in the 
Southern Hemisphere and give the possibility of building a global map of the CR sky.

In EW1 we have already used the preliminary data from Super-Kamiokande. Their final 
data, as well as data from MILAGRO, ARGO-YBJ, although obtained at slightly lower 
energies with the median value of 6-10 TeV agree well with the IceCube data at 20 TeV. 
It gives the opportunity to claim that excess  and dip regions cover a very broad range
 of declinations at tens of TeV energies.     

Keeping in mind that the harmonic fit does not give a good description of the observed 
anisotropy, we, nevertheless, consider that the change in the energy dependence of the 
amplitude and phase, which occurs at energies above $\sim$0.1 PeV and was first 
noticed by the EASTOP collaboration \cite{EasT}, is confirmed by the latest 
measurements and therefore worthy of interpretation.

The change of the amplitude and phase which occurs as the primary energy approaches the
 PeV region, with the well known knee at 3-4 PeV, supports the astrophysical origin of 
the knee and allows us to discount the nuclear physics model, in which the knee is 
caused by a change of the interaction characteristics of the primary CR particles with 
air nuclei. 

Before continuing mention should be made of the remarkable 20 degree-radius excesses 
observed by MILAGRO and Tibet. Heliospheric phenomena are a distinct possibility, or 
very local ISM irregularities, although an explanation in terms of EAS initiated by 
gamma rays from distant discrete sources cannot be ruled out.

At this stage it is necessary to describe the objectives of the present work, 
specifically. They are as follows. \\  
(i) To interpret the $A, \phi$ values up to $logE,GeV$ = 5, i.e. why the amplitude and 
phase should be nearly energy independent and why the values are as measured. \\
(ii) To interpret the values beyond $logE,GeV$ = 5, in terms of a single source, or 
otherwise.
\section{Interpretation of the anisotropies below $logE = 5$}
\subsection{Initial examination of the observations and comparison with prediction}
It is evident from Figure 1a that there is consistency between the various sets of data
 - a welcome result in Cosmic Ray Physics. A mean line has a slow increase in $logA$ 
with increasing energy to a maximum at about $logE,GeV = 3.5$. A slow fall follows 
to a minimum at about $logE,GeV = 5.5$.

The actual degree of consistency for the amplitudes from one experiment to another has 
been examined in two ways: by studying the dispersion in amplitudes (logA) about the 
smooth line and by doing likewise in terms of the number of standard deviations from 
the line for each experimental data point. The 'smooth line' is simply a line through 
the unweighted averages of the values for half-decadal energy ranges with no overlap, 
it is not shown so as not to confuse the Figure 1, but it can be easily visualised. The
 result for the former is that the median displacement of the points from the line is 
$\Delta logA = 0.1$, with 5\% beyond $\Delta logA = 0.3$. For the standard deviations 
$\sigma$, the median is $0.08\sigma$, with 5\% beyond $3\sigma$, a somewhat higher 
percentage than expected for a Gaussian, but not dramatically so.

The former shows that the 'curvature' in logA vs logE, with a peak at about logE, GeV =
 3.5 higher than the mean of the values of logA at logE = 2.0 and 5.0 
by $\Delta logA = 0.6$, is genuine. The latter gives confirmation.

Turning to the phases, for absolute values the median dispersion is $\sim 10^\circ$ 
with 5\% greater than $60^\circ$. In terms of the number of standard deviations, the 
method cannot be used because in one third of the points errors are difficult to read
from the original figures. 
Nevertheless, the dispersion of the phase, as such, gives no cause for claiming other 
than a straight line for log(RA) vs logE: there is no large scale feature mirroring the
 curvature in logA. The significance of this result will be considered later.

Concerning our prediction (~thin lines in Figure 1a \cite{EW1}~), there is modest 
agreement between our mean and the observations to $logE,GeV \approx 3.5$ but an 
increasing discrepancy above. A change must be made to the model. Two possibilities are
 considered: \\
(i)  There is a Giant Galactic Halo \\
(ii) The CR lifetime at high energies falls down more slowly with rising energy 
than it does at low energies  
\subsection{The amplitude of the anisotropy}
\subsubsection{A Giant Galactic Halo}
 We start with a suggestion to invoke a Giant Galactic Halo (GGH) having the uniform 
spatial distribution at least within the Galactic Disc and diffusive properties 
different from those in the conventional '1kpc Halo'. Since the uniform spatial 
distribution means complete isotropy with $A = 0$ the contribution of the GGH to 
the total CR flux would help to understand the low value of the anisotropy. 

The presence of such 
a Halo has received important support very recently from the Chandra X-ray Observatory 
\cite{Gupt}. The observations indicate a Giant Halo of radius over 100kpc, mean 
temperature $(1-2.5)\cdot10^6$$^\circ$K and total mass of $\sim 10^{11}M_\odot$. 
The energy density of the plasma is thus 0.15 to 0.4 eVcm$^{-3}$, close to that in CR 
in the Galactic Disc.

Empirically, the average of the predicted $A(E)$ samples in Figure 1a exceeds the   
observed amplitudes at $logE,GeV \approx 5.0 - 5.5$ by a factor of about 20. 
 This would mean that at these energies some 95\% of the CR come from the GGH. 
If the CR energy spectrum in the GGH is steeper than that in 
the Galacti Disc then at smaller energies the fraction of CR from the Disc would be 
smaller still. 

The cause of the steeper energy spectrum of particles in the GGH could, in principle, 
be the existence of the Galactic Wind which carries away low energy particles from 
the Galactic Disc to the GGH. However, simple calculations of the particle balance
in the system of Galactic Disc, Giant Galactic Halo and Extragalactic Space shows 
that if this system is in dynamic equilibrium, the GGH is not affected by the
Galactic Wind. The preservation of the particle balance means that the particle input
to the Disc from SN explosions has to be equal to the particle output from the Halo
into Extragalactic Space. The equations are similar to those used in the 
leaky box model:
\begin{equation}      
N^d(E)/\tau^d(E) + N^d(E)f/\tau^w(E) = Q(E) = N^h(E)/\tau^h(E)
\end{equation}
Here, $N^d(E), N^h(E)$ are energy spectra of particles in the Disc and Halo, 
$\tau^d(E), \tau^w(E), \tau^h(E)$ are lifetimes of particles in the Disc, Wind and Halo
 respectively and $Q(E)$ is the spectrum of particles injected by SNR into the Disc.
The second term in the equation takes into account the output spectrum of particles
carried away by the Wind. It is presented as the fraction of the Disc particles $f$ 
and the energy dependence of this fraction is included into the lifetime $\tau^w(E)$.

If, for eample, all energy dependencies in (1) are of the power law type, ie
\begin{equation}
Q(E)=Q_0(\frac{E}{E_0})^{-\gamma_{inj}}; \tau^{d,w,h}(E)=\tau^{d,w,h}_0(\frac{E}{E_0})^{-\delta^{d,w,h}}; 
\end{equation}   
then the solutions of the equations (1) are
\begin{equation}
N^d(E)=\frac{Q_0\tau^d_0(E/E_0)^{-(\gamma_{inj}+\delta^d)}}{1+\frac{f\tau^d_0}{\tau^w_0}(E/E_0)^{-(\delta^d-\delta^w)}} 
\end{equation}
\begin{equation}  
N^h(E)=Q_0\tau^h_0(E/E_0)^{-(\gamma_{inj}+\delta^h)}
\end{equation}
As an example we present the solution of equations (3) and (4) in Figure 2 for 
the numerical values shown in Table 2:
\begin{center}
\begin{tabular}{|c|c|c|} \hline
Object         & $\tau_0$,year & $\delta$   \\ \hline
Halo           & $4\cdot10^9$    & 0.6      \\
Disc           & $4\cdot10^7$  &   0.33     \\
Wind           & $4\cdot10^6$  &   0.2      \\ \hline
$\gamma_{inj}$ & \multicolumn{2}{c|}{2.4}   \\
$E_0,GeV$      & \multicolumn{2}{c|}{1.0}    \\
f              & \multicolumn{2}{c|}{0.5}   \\ \hline
\end{tabular}
\end{center}
{\footnotesize Table 2. Numerical values of the parameters used, as an example, 
in the calculations shown in Fig.2.}
\begin{center}
\begin{figure}[hbt]
\includegraphics[height=15cm,width=6cm,angle=-90]{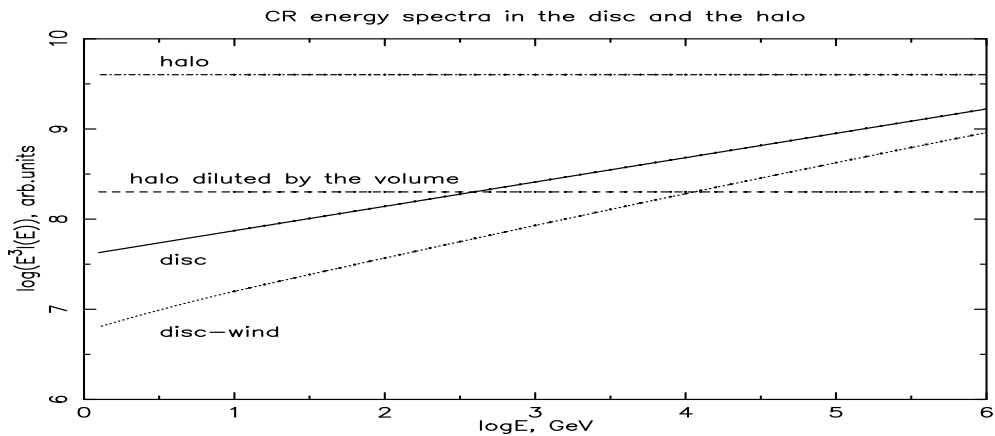}
\caption{\footnotesize An example of the calculations illustrating the possible 
contribution of the Halo to the CR intensity: full line - Disc with no wind, dotted
 line - Disc with the wind, dashed line - Halo, dash-dotted line - CR intensity in the
halo is diluted by the 20-fold larger volume.} 
 \label{fig:fig2}
\end{figure}
\end{center}
 It is seen that even for such a reasonable set of parameters shown in Table 2 and for 
the total number of particles in the Disc and the Halo it is impossible to reach 
a 20-fold excess of Halo over the Disc. If we take into account the large volume of the
 Halo (~in Figure 2 it was assumed the cylindrical shape of the Halo with the tenfold 
increase of the vertical scale height~) then the contribution of the GGH to the CR
{\em intensity} (~not the number of particles~) at PeV energies will be only 10-20\%.
The wind cannot amplify the contribution of the Halo and reduce the amplitude of the 
anisotropy.     
\subsubsection{A non-standard propagation model}
The amplitude of the anisotropy $A$ is connected with the characteristics of the CR 
density $N$, its gradient $gradN$ and the diffusion coefficient $D$ as 
$A = \frac{3DgradN}{cN}$, where $c$ is the
speed of light. In our propagation model $D(E) = H^2_d/\tau_d(E)$, where $H_d$ is the
 vertical scale height of the Disc, which we assumed to be 1kpc and $\tau_d(E)$ is 
defined above as the lifetime of the CR particles in the Disc. Therefore,
\begin{equation}
A = \frac{3H^2_dgradN}{cN\tau_d(E)}
\end{equation}

This expression shows that there is a unique relationship between the mean CR lifetime 
$\tau_d(E)$ and the anisotropy $A$. Figure 3 shows the lifetime implied by our model
(~which has $\langle \tau_d \rangle = 4\cdot 10^7E^{-0.5}$y with $E$ in GeV~) as
given in Figure 1a with $\delta = 0.5$. Elsewhere, we have made predictions of 
the energy spectra using $\delta = 0.33$ \cite{EW3} and others, eg \cite{Bla1} have 
made calculations for the same value of $\delta = 0.33$. Figure 3 shows three variants 
of the standard model.
\begin{figure}[hpbt]
\includegraphics[height=15cm,width=6cm,angle=-90]{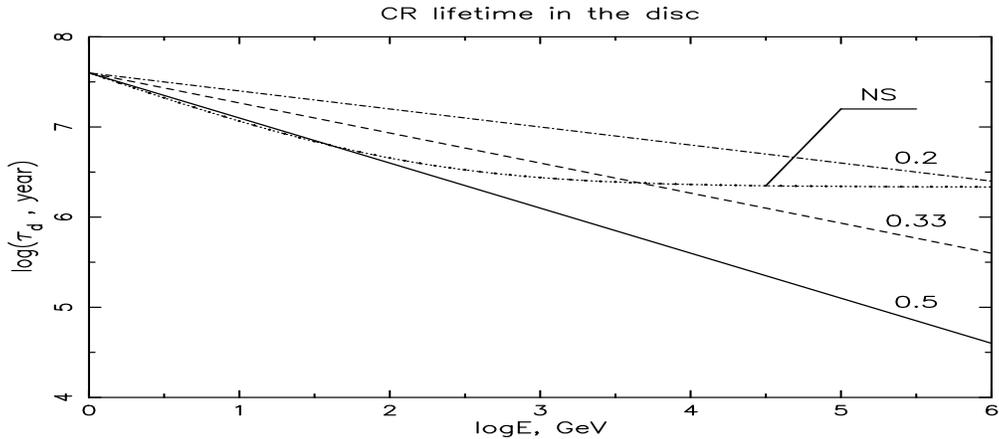}
\caption{\footnotesize Lifetime $\tau_d$ vs Energy (~protons, only, assumed~) 'NS'is 
for non-standard propagation model needed to explain the observed anisotropy - the 
first harmonic amplitudes. Numbers at the other lines are values of $\delta$ - the 
exponent in the usual expression  for mean lifetime vs energy: 
$\tau_d \propto E^{-\delta}$ }
\label{fig:fig3}
\end{figure}

Also shown, as 'NS', is the mean lifetime versus energy needed to give a best smooth 
line going through the experimental points in Figure 1a. It is seen that a dramatic 
flattening is needed above about logE,GeV = 3. In fact, when allowance is made for a 
modest Halo \cite{EW1} and downward fluctuations at the 10\% probability level, 
$\delta = 0.2$ would suffice.

Consequences of the Non-Standard (NS) model are considered in some detail in the 
Appendix, but some general remarks are given here.

It has long been known that $\delta = 0.6$ is needed for $E \leq 100$ GeV/nucleon
in order to explain the secondary to primary ratio (S/P) as a function of energy
\cite{Ober}. At higher energies smaller values of $\delta$ (~0.3 to 0.5~) have been 
commonplace. Indeed, in our own work \cite{Erl1} involving 'anomalous diffusion' values
 in the range $0.25 < \delta < 1.0$ were considered. There is thus a precedent for 
considering small values of $\delta$. 

Figure 4 shows the anisotropy amplitude logA vs LogE for 10 samples of the SNR 
space-time distributions calculated with the 'non-standard' (NS) lifetime of CR protons
shown in Figure 3. It is seen that the rise of the amplitude with the energy seen in 
Figure 1a changed to a nearly constant value at rather low absolute level. The change 
of the gradient $\frac{gradN}{N}$ expected with the change of the diffusion coefficient
does not compensate the reduction of the amplitude due to the change of the lifetime.
\begin{center}
\begin{figure}[hbt]
\includegraphics[height=15cm,width=6cm,angle=-90]{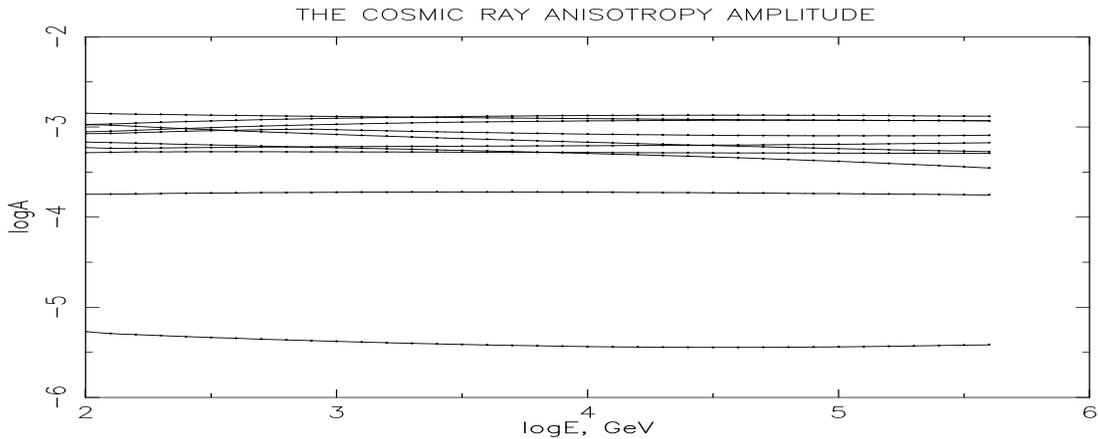}
\caption{\footnotesize The anisotropy amplitude for 10 samples of SNR with the  
'non-standard' propagation model.} 
\label{fig:fig4}
\end{figure}
\end{center}
\subsection{The phase of the anisotropy}
Again, there is a reassuring consistency of the phases from the different experiments
 (~Figure 1b~). A slow change of phase is indicated, from a phase of $90^\circ$ at 
logE,GeV = 2 to about zero at logE,GeV = 5.

As might be expected, at 'low energies' where the local conditions of ISM are 
relevant, the experiments indicate a more complicated anisotropy pattern than a simple 
first harmonic and this aspect is taken up in the next section.
 \subsubsection{Two-dimensional maps}
    The variation of the CR flux with the right ascension is better not analysed in 
terms of the simplified one-dimensional picture with the phase, often integrated over a
 large range of declinations. Rather one should examine the two-dimensional maps, 
provided by the publications, which often show a more complicated picture with a few 
excesses in different parts of the sky at different RA and $\delta$. We are mindful of 
the potential criticisms of Andreev et al., 2008 \cite{Andr} who criticise the validity
 of two-dimensional maps. Nevertheless, we continue to use the results of the 
experimenters themselves. 

Figure 5 shows a
 schematic view of the location of CR excesses in both Equatorial and Galactic 
coordinates for the 5 new experiments, which published these maps: MILAGRO. ARGO-YBJ, 
Super-Kamiokande, IceCube and IceTop. 
\begin{figure}[h]
\includegraphics[height=15cm,width=10cm,angle=-90]{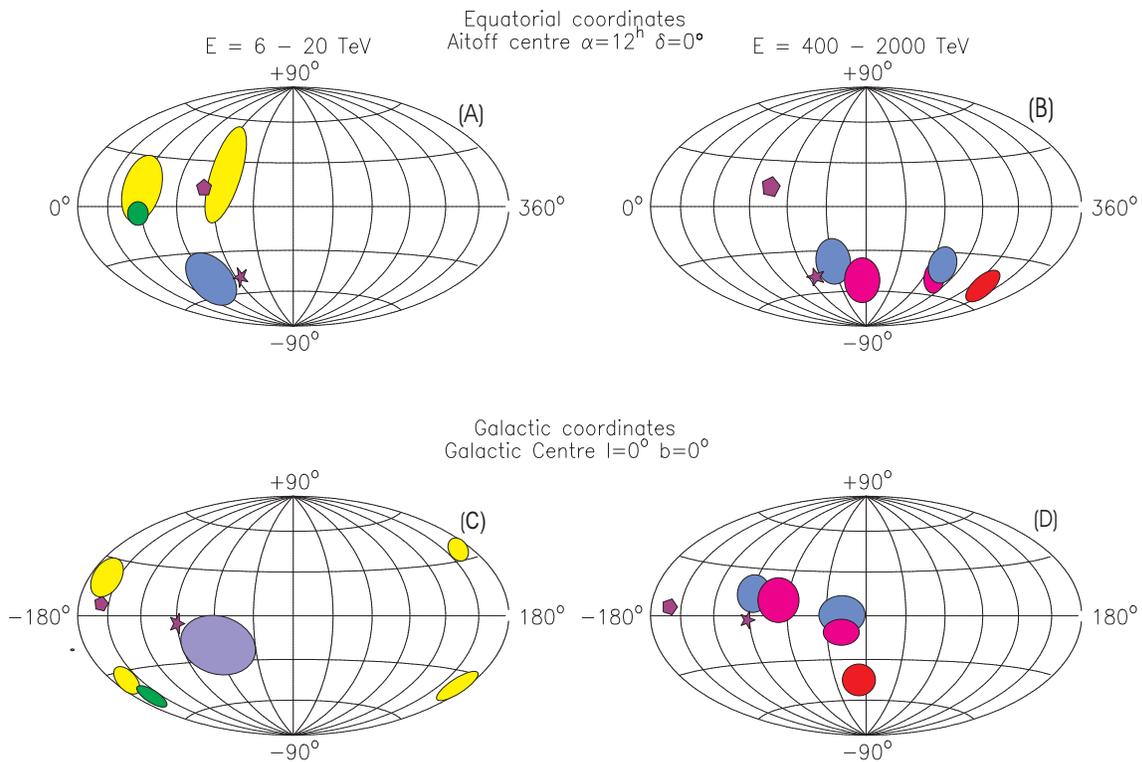}
\caption{\footnotesize Schematic maps of CR excesses in Equatorial (a,b) and Galactic 
(c,d) coordinates for two energy domains: 6 - 20 TeV (a,c) and 400 - 2000 TeV (b,d).
Excess regions observed in MILAGRO and ARGO experiments nearly overlap and are shown in
 yellow. Super-Kamiokande - green, IceCube - blue, IceTop - magenta for 400 TeV CR, 
red for 2000 TeV. The dark pentagon indicates the position of the Monogem Ring SNR, 
the dark star - the position of Vela X. } 
\label{fig:fig5}
\end{figure}

Positions of CR excesses observed in these 
experiments often split and overlap and we decided to give very schematic presentation 
of these excesses just to illustrate the location of 
the excesses in the CR sky. Contours of the ellipses in Figure 5 embrace very 
approximately the areas of excesses exceeding the mean background by three standard
 deviations. The upper half of Figure 5, panels (a) and (b)
show maps in Equatorial coordinates, the lower half, panels (c) and (d) show the same
maps in Galactic coordinates.
Panels (a) and (c) show the excesses observed in the lower, 6 - 20 TeV energy domain.
The overlap of the MILAGRO, ARGO-YBJ and Super-Kamiokande gives a measure of 
confidence for these measurements and confirms the puzzling conclusion made in 
\cite{EW1} that in spite of the fact that the bulk of SNR, pulsars and other 
potential CR sources are in the Inner Galaxy surrounding the Galactic Centre, the 
excess of CR is observed in the opposite, Anti-Centre direction. The new point in these
 measurements is actually connected with the fact that MILAGRO and ARGO-YBJ 
collaborations observe two excesses in the same direction of the Anti-Centre, but both 
in the Northern and in the Southern Galactic Hemisphere. This gives an argument in 
favour of the conclusion that the CR at TeV energies originate in sources whose 
directions span a large range of Galactic latitudes.

The measurements of IceCube at the higher energy of $\sim$20 TeV show that presumably 
the excess region shifts in the direction of the third and fourth quadrants.    

It would be expected that at tens of TeV energies the giroradius of particles in the 
typical magnetic fields of a few microgauss should be very small, $\sim$0.01pc. The 
diffusion of particles is thus strongly dependent on very local phenomena: the 
direction
 of the local regular magnetic field component, 'modulated' by the random field and the
 near-absence of a gradient of the CR density. The latter is determined by the 
influence of the 'Local Fluff' and the spiral Orion Arm which are located in the 
direction of the Outer Galaxy and can overcome the influence of the background CR flux 
from the sources in the Inner Galaxy \cite{Pare}. The Local Fluff is a well known and 
important feature of the local ISM (~see reference \cite{Pare} for the original 
reference; much later work has confirmed the identification~). It is of about 
10 pc extent and can clearly affect the propagation  of low energy CR, i.e. those of 
scattering length of this order and less. The effect will be significant in view of the
 fact that there will be transit time and, thereby, intensity differences from nearby 
sources depending on direction.  

Inspection of Figures 1b and 5a,c shows that the value of $\phi$ is close to that 
expected for an excess of CR intensity from along the $B$-direction over the 'local' 
$B$-direction, the best fit being with the very local $B$-direction but the 
uncertainty in the actual values (typically $\pm 20^{\circ}$), and the likely effect of
 local field irregularities make the discrepancy understandable. Another important 
aspect is the gradient of the CR energy density that would be expected in the absence 
of regular field effects. The near constancy of amplitude and phase indicates that the 
'true' anisotropy (i.e. that which would occur in the absence of a regular field 
component ) is very small. 

The direction to the Galactic Anti-Centre from which the excess of CR arrive, as 
has been already remarked, is presumably due to factors connected with the small 
Galactocentric gradient of CR emissivity \cite{EW4}. In that work it was suggested that
 shear in the ISM in the relatively undisturbed parts of the Outer Galaxy ISM was 
responsible for the (near-) lack of gradient. This would be in addition to the random 
nature of the SNR sources.

A further factor is that the reentrant PeV particle from the GGH will, themselves, have
 a small anisotropy from the general direction of the Galactic Centre.

\section{Interpretation of the anisotropies above $logE = 5$}
Figure 1 and panels (b,d) of Figure 5 show results of anisotropy measurements at the 
relatively high energies of about 400 - 2000 TeV. They were made by two
independent detectors: IceCube and IceTop at the South Pole. It is seen that the 
amplitude of the anisotropy, after falling down to the minimum at several hundred TeV, 
 started to rise again above $\sim$400 TeV, exceeding the value of $A \approx 10^{-3}$
in the PeV region.

The phase of the first harmonic dropped by nearly $150^\circ - 180^\circ$.  An 
inspection of the two-dimensional map shows that the results of two measurements at the
 comparable energy of 400 TeV
partly overlap, giving support to each other. The excesses observed at this energy 
occupy the region close to Galactic Disc in the fourth quadrant. At the highest energy 
of 2000 TeV IceTop experiment observed that the excess moved to the direction still 
closer to the Galactic Centre, but shifted to higher Southern Galactic latitudes.   

At PeV energies the scattering length approaches 100pc and it would be expected that 
the influence of the local environment would become smaller and more distant regions of
 the Galaxy enter the scene. At larger distances from the solar system the higher 
density of sources in the Inner Galaxy overcomes the influence of the local environment
and the CR excess moves toward the direction of the Galactic Centre. However, the 
existence of the knee and other fine structures in the CR energy spectrum give  
evidence that non-uniformity in the space-time distribution of sources still 
reveals itself at PeV energies and above.     

Figure 5 shows the positions for the potential sources: the Monogem Ring SNR and Vela X
 pulsar with its pulsar wind nebula. The latter is young and can be a potential source 
if the particles can escape from the SNR associated with them. It is seen that both 
sources are in the Outer Galaxy and close to the Galactic Disc. The Monogem Ring SNR is
 nearby
one of the excess regions indicated by MILAGRO and ARGO-YBJ at lower 6-20 TeV energies.
It is not far from the direction towards the Anti-Centre and can in principle   
contribute to this excess. Counterargument that the excess area does not cover the 
position of the Monogem Ring SNR centre can be overcome by the fact that this SNR has a
rather large size with a ring diameter of about $18^\circ$ \cite{Thor}. So we cannot 
reject the possible contribution of the Monogem Ring SNR to the anisotropy at energies 
below 0.1 PeV on the basis of just anisotropy arguments. On the other hand, if the 
Monogem Ring is responsible for the knee and has rather flat energy spectrum, its 
contribution at energies below 0.1 PeV is negligible. Another worry is that this SNR 
cannot be associated so far with any excess at the higher 400-2000 PeV energies, 
although it has to be remarked that all the indicated measurements in panel (b) of the 
Figure 5 have been made at the South Pole and do not cover the positive declinations 
where the Monogem Ring is situated.   

Vela X can in principle have relevance to the formation of the excess observed by 
the IceCube experiment in both energy domains. However, it is not in the direction
of the Galactic Centre, but just close to the border between the third and fourth 
quadrants.  

As for its characteristics as the potential CR source, not only is there the problem of
 particle escape from the encompassing SNR but the time window for emission of 
particles of adequate energy is not clear. This lack of clarity arises because of 
uncertainty about the pulsar's period at birth $P_0$, its' past rate of energy loss and
 the efficiency for CR acceleration. A literature survey yields a range of $P_0$: 18 to
 near 89ms. For most of the range, however, there should be an adequate window for PeV 
particles to be generated and the diffusion time from pulsar to the Earth is 
acceptable.  The remaining questions relate to CR efficiency and particle escape.
\section{Comparison with other work}
Since the anisotropy is very small the work in this field is difficult because it 
requires high statistical accuracy. That is why the experimental data on the anisotropy
 at sub-PeV and PeV energies are rather poor. The KASCADE experiment did not find any 
anisotropy \cite{Anto} and we give their upper limits in this and our previous 
paper \cite{EW1}. 

Updated EAS-TOP data give only the amplitude and phase of the first 
harmonic \cite{EasT} and not two-dimensional maps of the excesses in the CR flux. We 
used their data in our Figure 1. 

In addition to our earlier work on the single source others have contributed to the 
development of the idea. Mention has already been made (in \S1) of the work of  
Sveshnikova et al. \cite{Svesh} in which both known and simulated nearby SNR were 
considered. The dip in Figure 1a was not reproduced. Vela X was discounted because of 
the wrong phases above $logE,GeV$ = 5 appear not to have been included. Nevertheless, 
the work mentioned adds to the attention given to the role of the anisotropy as well as
 CR energy spectrum in defining a (possible) single source.
Other workers who have developed our stochastic source distribution model include 
\cite{Bla1} and \cite{Pohl}. The idea of a 'non-standard model' was not mentioned, 
however.

Mention should also be made of very recent work \cite{Desi}. The authors have claimed 
that heliospheric effect may be responsible for part at least of the anisotropy to 
energies as high as 100 TeV. This is a remarkable result insofar as heliospheric 
effects are usually considered to be small by 1 TeV. Confirmation, or otherwise, is 
awaited. In any event, our conclusions, and those of others, should not be affected 
beyond $logE,GeV = 5$ where the case for the effect of a single nearby source is made.
\section{Conclusions}
Up to $logE,GeV \sim 5$ there is consistency in the measurements of large scale (first
 harmonic) anisotropies and near constancy of phase: an important result enabling 
detailed analysis to be worthwhile. Transport along the 'local' 
magnetic field lines seems a likely cause although the direction of flow - from the 
general direction of the Galactic Anti-Centre - is surprising. The random space-time
configuration of CR sources with the contribution of a few sources such as the Monogem 
Ring and Vela X SNR can play a role in the formation of the excess CR flux from the 
general direction of the Galactic Anti-Centre. A Giant Galactic Halo is inadequate to 
account 
for the comparatively small measured anisotropy. Another discrepancy to be explained is
 the lack of convexity in logA vs logE plot in our 'conventional model'. Instead, a new
 model is invoked: a 'Non-Standard' propagation model. Its acceptance, or otherwise, 
awaits further analysis.
   
Contunuing with 'low' energies (~$logE,GeV \leq 5$~), there is no obvious rapid change 
of the anisotropy with logE, as yet. The 'New Component', claimed by us \cite{EW5} from
 observations of 
the detailed spectral shapes of many elements might be expected to give a change of 
anisotropy amplitude/phase at 100 GeV. Unfortunately, this is just the lower limit for 
our study, and, in any event, solar modulation causes distortions below about 100 GeV. 
There are no features in the energy range logE,GeV: 2 to 5 which can be attributed to 
fine structure (~although there is an interesting hint of a decadal periodicity of 
phases with respect to logE~).
  
At higher energies there is evidence for a change in anisotropy amplitude and phase, 
although the experimental data are, as yet, rather sparse and cover mostly the area of 
negative declinations. They indicate that the amplitude of the first harmonic begins to
 rise with the energy and the phase changes to the opposite. The CR flux excess in  
Galactic coordinates shifts to the fourth quadrant and at the highest PeV energies to 
the direction of the Galactic Centre. An explanation in terms of
 the onset of a significant contribution from the sources in the Inner Galaxy and some 
of the local sources is promising. Although the Monogem Ring SNR is still possible, 
Vela X is a distinct possibility if its birth period were short enough 
($P_0 \simeq 60$ms) and the particles were able to penetrate the surrounding Vela SNR 
envelope.

{\bf Acknowledgements}

The authors thank the Kohn Foundation for supporting this work.

\vspace{5mm}

{\bf Appendix. A non-standard propagation model}

It is shown in Figure 3 that the form of the $log(\tau_d)$ vs logE dependence is 
non-standard, in the sense that it is not simple power law, in contrast with the usual 
assumption. In fact, as remarked already, the assumption of a unique, energy 
independent form of $\delta$ is simplistic: propagation of CR through the ISM is, no 
 doubt, a process of great complexity.

In \cite{Erl1} varying degrees of turbulence were considered, giving rise, in turn, to 
varying degrees of anomalous diffusion (AD). Our adopted form of AD was characterised 
by the parameter $\alpha$: 0.5 to 2.0, corresponding to $\delta = \frac{\alpha}{2}$ 
from 0.25 to 1.0.
The lower the value, the greater the degree the degree of AD. Thus, Figure 3 indicates 
the need for anomalous diffusion becoming more and more important as the CR energy 
increases. 

The cause of this increase in anomalous diffusion is not immediately apparent. It would
 appear that higher energy particles sample parts of the Inner Galaxy (~where the 
turbulence is bigger~) to a greater degree than those of lower energy. In turn, this 
suggests that the scale height increases with energy, in distinction to the usual 
assumption of energy independence. The solution may lie in the model in which particles
 spiral around field lines and transfer from one line to another before joining a line
that leads to escape from the Galaxy. The transfer probability may be energy dependent.
 The conventional diffusion model itself is not without problems, in that CR are 
assumed to diffuse until they reach a height where they immediately escape; such a 
model is, again, simplistic. The conventional Galactic Wind may have a role to play, in
 that its effect is energy dependent. However, at TeV energies and above its effect 
would not be expected to significant.

Another aspect of the Galactic Wind may be relevant at energies approaching on PeV, 
however. This relates to the possibility of acceleration of particles at the 
termination shock, following \cite{Volk}. Insofar as the mechanism sends energetic CR 
back towards the Galactic Disc the termination shock acts as 'a reflecting boundary'. 
This is just what is required in our NS model, although it must be pointed out that the
 energies considered in \cite{Volk} are somewhat higher than those considered here.

Understandably, changing the lifetime of CR requires a change in the CR injection 
spectrum. Specifically, the injection spectrum will need to be steeper. We have adopted
an exponent (~$\gamma_{inj}$ in $N(E)dE = CE^{-\gamma_{inj}}dE$~) equal to 2.15 and 
this fits the observed proton energy spectrum for $\delta = 0.5$. Changing $\delta$ to
 0.2, for example, would need $\gamma_{inj} = 2.45$, a value increasingly away from the
 standard $\gamma_{inj} = 2$ for Fermi acceleration in a SNR envelope.

However, help is at hand when consideration is given to the time sequence of 
acceleration in SNR. As we have shown \cite{EW6} the higher energy particles achieve 
most of their energy at later times. Specifically, most of the CR of $logE > 4$ are 
accelerated in the last 20\% of the time of the expansion of the remnant (~64 to 80 
kyears in our model~). Thus, a steepened injection spectrum would arise quite naturally
 if SNR were leaky at these late ages. In fact, inspection of the observed SNR shows 
that they are rarely spherical and leakage seems inevitable. This feature is 
particularly marked for those SNR in the Inner galaxy. A further fact of relevance is 
the inevitable spread in the maximum energies of CR from SNR, arising from the variety 
of total energies, ISM densities and magnetic fields.

Finally, it should be remarked that in this scenario the 'Single Source' will need to 
have been an energetic SNR which did not leak prematurely.

\end{document}